\documentclass[12pt]{iopart}

\usepackage{color}
\usepackage{graphicx}
\usepackage{subfigure}

\usepackage{cite}

\newcommand{\ket}[1]{\ensuremath{\left|#1\right>}}

\newcommand{\yb}{$^{171}$Yb$^+$ }

\newcommand{\sfz}{$^2$S$_{1/2}\ket{F\!=\!0}$ }
\newcommand{\sfzc}{$^2$S$_{1/2}\ket{F\!=\!0,\,m_f\!=\!0}$ }
\newcommand{\sfo}{$^2$S$_{1/2}\ket{F\!=\!1}$ }

\newcommand{\sfoc}{$^2$S$_{1/2}\ket{F\!=\!1,\,m_f\!=\!0}$ }

\newcommand{\pfz}{$^2$P$_{1/2}\ket{F\!=\!0}$ }
\newcommand{\pfo}{$^2$P$_{1/2}\ket{F\!=\!1}$ }
\newcommand{\dfo}{$^2$D$_{3/2}\ket{F\!=\!1}$ }
\newcommand{\sfzn}{$^2$S$_{1/2}\ket{F\!=\!0}$ }
\newcommand{\sfonm}{$^2$S$_{1/2}\ket{F\!=\!1}$ }

\begin{document}


\title{Single qubit manipulation in a microfabricated surface electrode ion trap}

\author{Emily Mount$^1$, So-Young Baek$^1$, Matthew Blain$^2$, Daniel Stick$^2$, Daniel Gaultney$^1$, Stephen Crain$^1$, Rachel Noek$^1$,  Taehyun Kim$^1$\footnote{Present address: Quantum Technology Lab, SK Telecom, Seongnam-si, Gyeonggi-do, Korea 463-784}, Peter Maunz$^1$\footnote{Present address:
Sandia National Laboratories, Albuquerque, New Mexico, USA} and Jungsang Kim$^1$}

\address{$^1$Fitzpatrick Institute for Photonics, Electrical and Computer
Engineering, Duke University, Durham, North Carolina, USA}
\address{$^2$Sandia National Laboratories, Albuquerque, New Mexico, USA}

\ead{jungsang@duke.edu}
\maketitle


\begin{abstract}
We trap individual \yb ions in a surface trap microfabricated on a silicon substrate, and demonstrate a complete set of high fidelity single qubit operations for the hyperfine qubit.  Trapping times exceeding 20 minutes without laser cooling, and heating rates as low as 0.8 quanta/ms indicate stable trapping conditions in these microtraps.  A coherence time of more than one second, high fidelity qubit state detection and single qubit rotations are demonstrated. The observation of low heating rates and demonstration of high quality single qubit gates at room temperature are critical steps towards scalable quantum information processing in microfabricated surface traps.

\end{abstract}

\section{Introduction}

Quantum information processing allows one to solve certain types of computational problems that are intractable for classical computers. Trapped ion systems are promising candidates due to their long coherence times, isolation from the environment and ease of manipulation.  For these reasons many basic quantum algorithms have been implemented with small numbers of ions in macroscopic ion traps, such as the Deutsch-Jozsa algorithm\cite{Gulde2003}, a Toffoli gate \cite{Monz2009} and deterministic teleportation \cite{Barrett2004, Riebe2004}.  

A functional quantum computer architecture requires that many qubits can be initialized, manipulated, and measured with high fidelity \cite{DiVincenzo2000}. In addition, efficient qubit transfer within the processor is also necessary \cite{Oskin2003}. Physical architectures that allow for two qubit gates between distant ions can provide a substantial advantage in efficient execution of useful quantum circuits \cite{Duan2004, Monroe2012}. However, the realization of such architectures is challenging and requires new technologies and a substantial integration effort \cite{Monroe2013}.

Traditionally, ion trapping experiments have been carried out in macroscopic 3-dimensional trap structures \cite{Benhelm2008}.  However, these traps would be difficult to scale to the large number of ions and functions required by a quantum processor.  In contrast, surface traps leverage modern microfabrication technologies to confine ions above a two-dimensional planar electrode structure. They provide a scalable platform for ion trap quantum computing \cite{Chiaverini2005,Kim2005a} by allowing complex trap structures that can contain multiple activity zones \cite{Chiaverini2005}, and closely-packed ion chains between which qubits can be transported by ion shuttling \cite{Kielpinski2002, Kim2005a, Steane2007}.  Substantial progress has been made on the trapping and shuttling of ions in microfabricated surface traps with a large number of trap segments \cite{Amini2010,VanDevender2010,Wang2010,Merrill2011, Moehring2011}.  While many groups have successfully fabricated and trapped ions in surface traps, qubit manipulation demonstrations have been limited to cryogenic experiments \cite{Wang2010, Brown2011, Leibrandt2009},  or experiments performed using microwaves \cite{Ospelkaus2011, Allcock2013}.

Here, we report the manipulation of a \yb hyperfine qubit in a microfabricated surface ion trap system operating at room temperature.  We show single qubit Rabi oscillations driven by both microwaves \cite{Timoney2008, Johanning2009, Webster2013} and Raman transitions using optical frequency combs. Additionally, we demonstrate control of the ion's motion using two nearly perpendicular beams of optical frequency combs \cite{Hayes2010}.  With adequate control of the ion motion we cool and measure the ion heating rate relevant for implementation of a two qubit gate such as the M{\o}lmer-S{\o}rensen gate \cite{Molmer1999}.

\section{Experimental setup}
\subsection{Trap}
\label{sec:Trap}

\begin{figure}[ht]
\centering
	\subfigure[]{
		\includegraphics[trim=0cm 0cm 0cm 0cm, width=.45\textwidth]
		{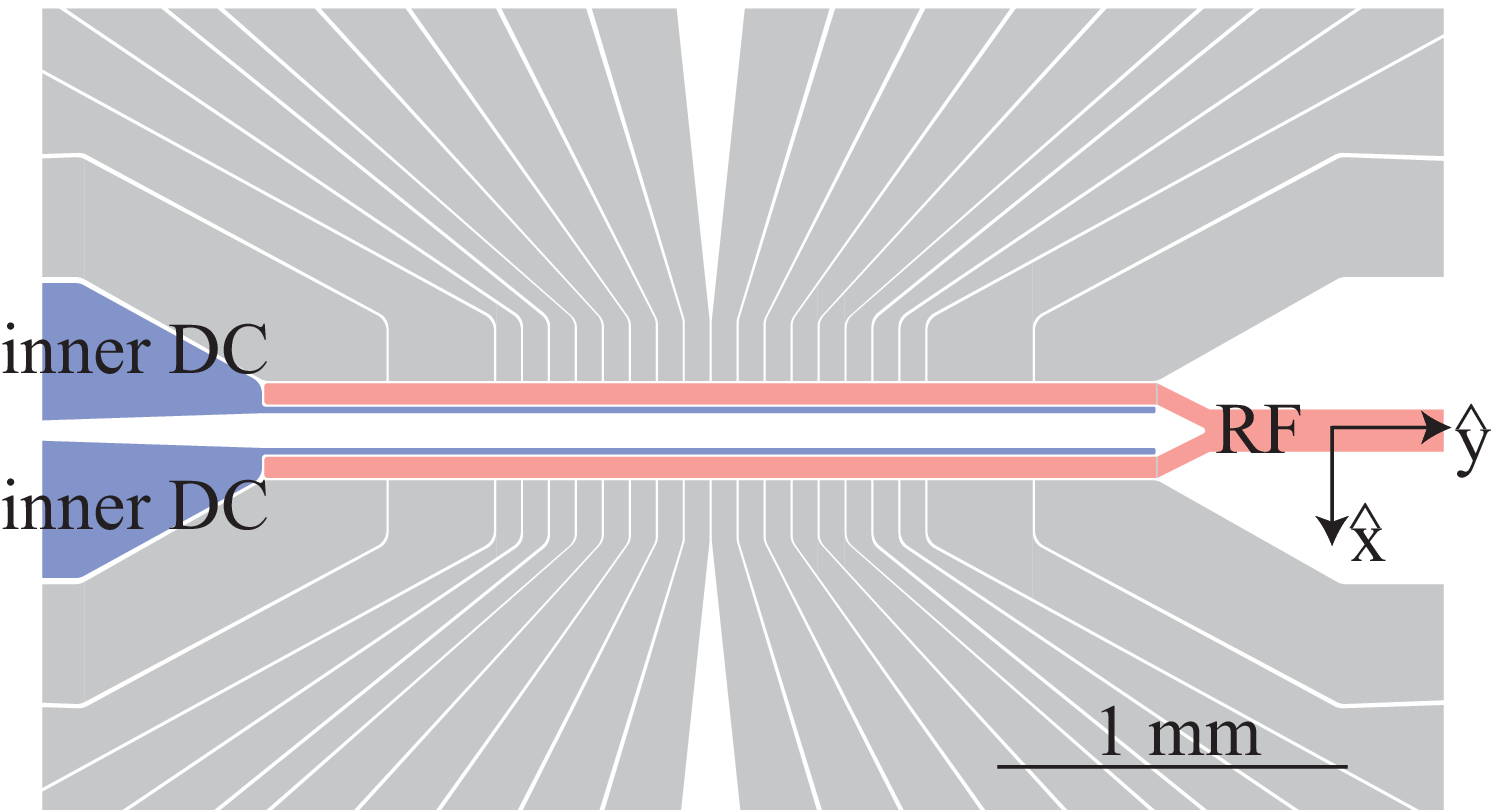}
		\label{fig:tbirdSchematic}}
		\quad
	\subfigure[]{
		\includegraphics[trim=0cm 0cm 0cm 0cm, width=.45\textwidth] 
		{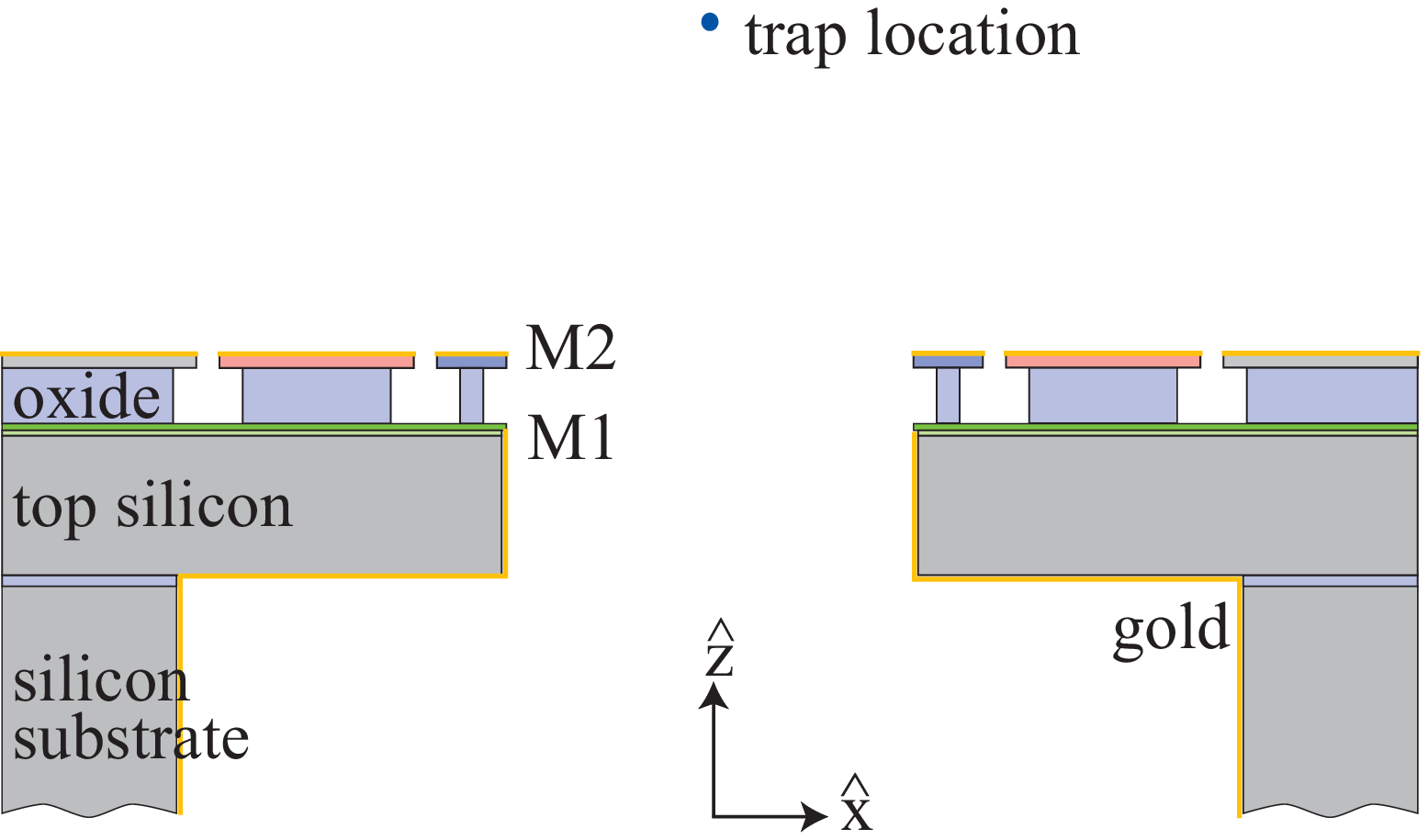}
		\label{fig:tbirdCrossSection}}
		\quad
\caption{(color online) Sandia Thunderbird trap.  (a) Schematic of the electrode configuration: The trap is a symmetric 6-rail design consisting of dual inner DC electrodes (blue), RF electrodes (red) and segmented outer DC electrodes (gray). The inner DC electrodes allow for rotation of the principal axes of the generated trapping potential. Located outside of the inner DC electrodes are 60\,$\mu$m wide RF electrodes.  (b) Cross-sectional diagram of the trap. The silicon substrate is shielded from RF fields by the first grounded metal layer (M1).  The second metal layer (M2) carries all of the trapping voltages and is separated from M1 by a $\sim$\,14\,$\mu$m thick oxide layer which, compared to a thinner oxide layer, reduces the capacitance between the RF electrodes and ground and increases the RF electrode breakdown voltage. Gold is evaporated from the top and bottom of the trap to minimize the amount of dielectric the ion is exposed to.}
\label{fig:sandiaFigure}
\end{figure}

Scaling trapped ion systems to a large number of qubits necessitates the use of complex multi-segmented trap structures. Fabrication of trap structures with a large number of electrodes is feasible using established microfabrication techniques based on photolithography \cite{Chiaverini2005}. While surface electrode traps have been made in both single metal layer designs \cite{Seidelin2006, Brown2007,  Allcock2010} and designs with multiple metal layers\cite{Stick2010, Doret2012}, the multiple metal layers are essential to connect electrodes to input/output (I/O) bond-pads when designing complex traps with multiple trapping zones and junctions.

The radio frequency (RF) surface electrode Paul trap used in this work features two metal layers separated by an insulating layer \cite{Stick2010, Allcock2011}. The geometry of the trap is a symmetric 6-rail design (shown in figure \ref{fig:tbirdSchematic}) with a 100\,$\mu$m wide slot etched through the substrate to allow for backside ion loading and optical access. Adjacent to the slot are split central control electrodes, two electrodes with RF voltages applied, and 40 outer segmented control electrodes.  The control electrodes have quasi-static voltages applied to them and are often designated as DC electrodes. The top metal layer is separated from the bottom metal layer by 14\,$\mu$m of oxide and each metal layer consists of 2.4\,$\mu$m of aluminum. The two RF rails have a width of 60\,$\mu$m and are separated by 140\,$\mu$m. Together they have a capacitance of $\sim$\,7\,pF to RF ground. The equilibrium trapping position is $\sim$\,80\,$\mu$m above the trap surface. The oxide insulating layer is controllably etched back so that the oxide is cleared away from the RF ground directly under the gap between the electrodes and the electrode metal overhangs the oxide, thus reducing the amount of insulator visible from the trapping region (figure \ref{fig:tbirdCrossSection}). This overhang makes it possible to evaporate a different metal on the top electrodes without causing shorts.  Two of the traps investigated had an additional 500\,nm gold layer evaporated onto the top surface.

The trap chip is mounted on top of a spacer in a 100-pin ceramic pin grid array (CPGA) package to facilitate lateral optical access. Wirebonds originate from the lower metal layer and feature low profile bonds protruding only a few tens of micrometers above the top metal plane. In addition, wirebonds are placed such that optical access is not obstructed.  On the package for one of the gold-coated traps, a filter capacitor of $\sim$\,1\,nF was installed between each DC electrode and ground to filter the residual pick-up of RF voltages, which is crucial for eliminating ion micromotion.  Each trap is electrically tested to ensure none of the electrodes are disconnected or shorted and thus fully functional.

\subsection{Vacuum system}

The stability of ions trapped in surface traps is sensitive to collisions with background gas molecules as trap depths are much lower than macroscopic 3-dimensional traps (typically less than a few hundred meV). The ion traps used in  these experiments are housed in a spherical octagon vacuum chamber to allow for ample laser beam access. The vacuum chamber contains 96 electrical feedthroughs for DC electrodes. Separate electrical feedthroughs are used for the RF voltages to minimize pick-up of the RF signal on the DC electrodes, which can cause ion micromotion. A thin sheet of aluminum with a 2.5\,mm wide viewing slit above the trapping zone is placed 3.8\,mm from the trap surface to shield the trapped ions from residual electric fields due to charge build up on the vacuum viewport.   For some experiments a non-evaporable getter (NEG) pump is placed 3.3\,mm away from the trap, between the trap and the solid aluminum ground shield, to reduce the vacuum pressure close to the ion. The vacuum chamber is constantly pumped by an ion pump, a titanium sublimation pump, and the non-evaporable getter pump. The pressure of the vacuum system was below what is measurable by the ion gauge used ($<$\,1.9$\times$10$^{-11}$\, Torr) for all experiments.

\subsection{Electronics}
\label{sec:Electronics}

The ion traps used in this experiment contain 42 DC electrodes. These electrodes are connected to a custom digital-to-analog converter (DAC) system outside of the vacuum chamber, used to provide the necessary DC voltages to the trap. The DC voltages are chosen to rotate the principal axes such that the cooling beam has a projection in each of the three principal axis directions, allowing for efficient Doppler cooling of all motional modes. RC filters with a $\sim$\,300\,kHz cutoff frequency are placed just outside the vacuum chamber, $\sim$\,330\,mm away from the chip.   Capacitive filters (900\,pF) are located on the CPGA package, close to the ion trap chip.  The RF trapping voltage is generated by a direct digital synthesizer (DDS), and supplied  through an RF amplifier (gain of 40\,dB) followed by a helical resonator (resonance at $\omega_{\mathrm{RF}}$\,=\,27.8\,MHz, $Q_{\mathrm{loaded}}$\,=\,280) \cite{Macalpine1959}.

\subsection{Qubit and optical setup}
\label{sec:qubit}

\begin{figure}[ht]
\centering
		\includegraphics[trim=0cm 0cm 0cm 0cm, width=.5\textwidth]
		{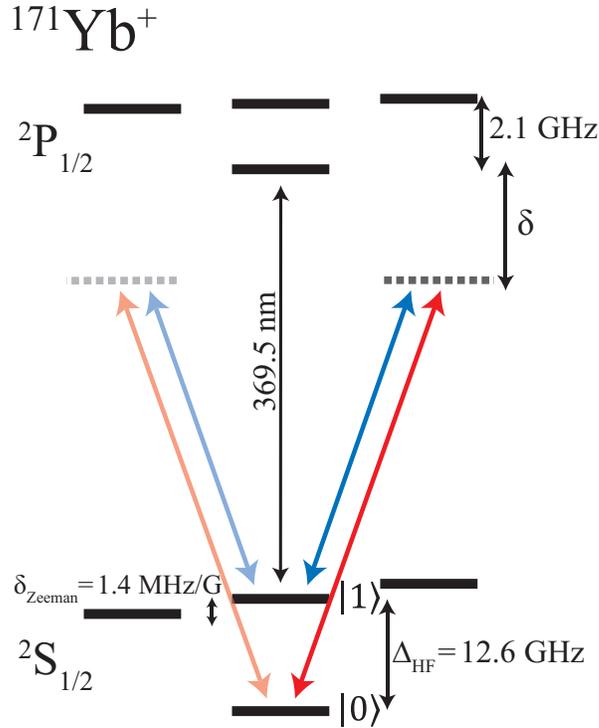}
\caption{(color online) Relevant \yb energy levels.  Co-propagating Raman beams (dark red and blue) can drive coherent transitions between qubit states. Raman beams with a $\Delta$k perpendicular to the trap axis (both dark and light red and blue) can also couple to the ion's transverse motional modes. }
\label{fig:RamanQubit}
\end{figure}

\begin{figure}[ht]
\centering
	\subfigure[]{
		\includegraphics[trim=0cm 0cm 0cm 0cm, width=.7\textwidth]
		{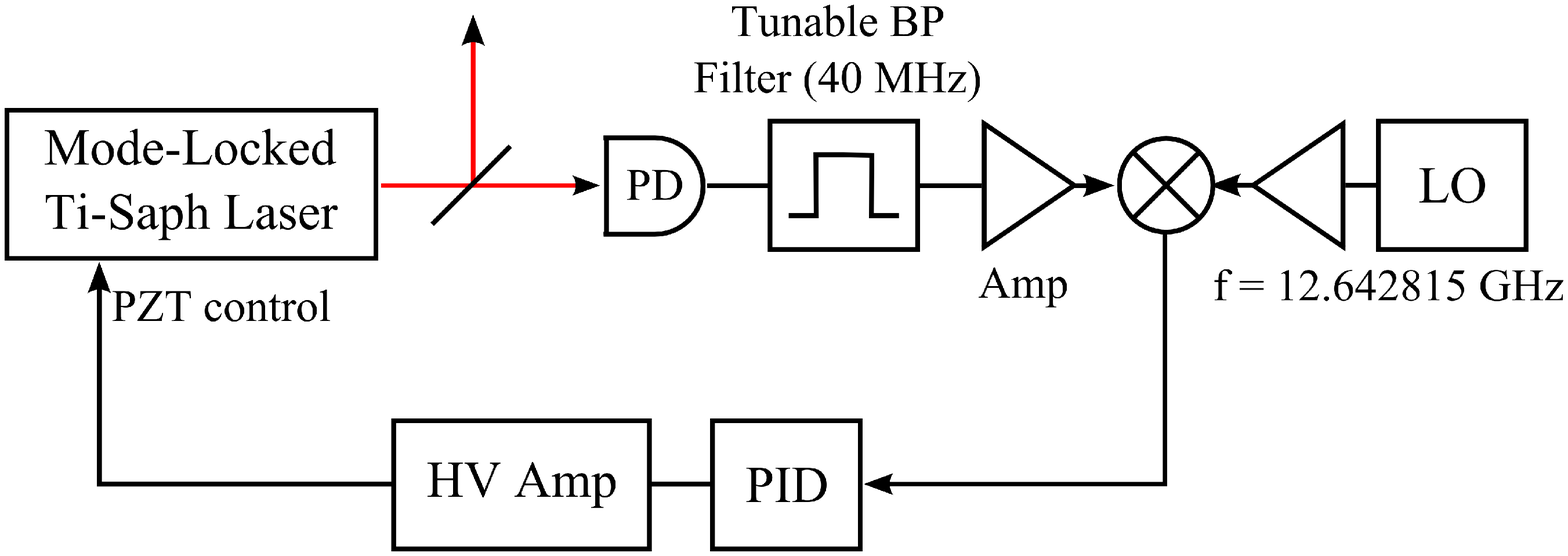}
		\label{fig:lockingfig}}
		\quad
	\subfigure[]{
		\includegraphics[trim=0cm 0cm 0cm 0cm, width=.6\textwidth] 
		{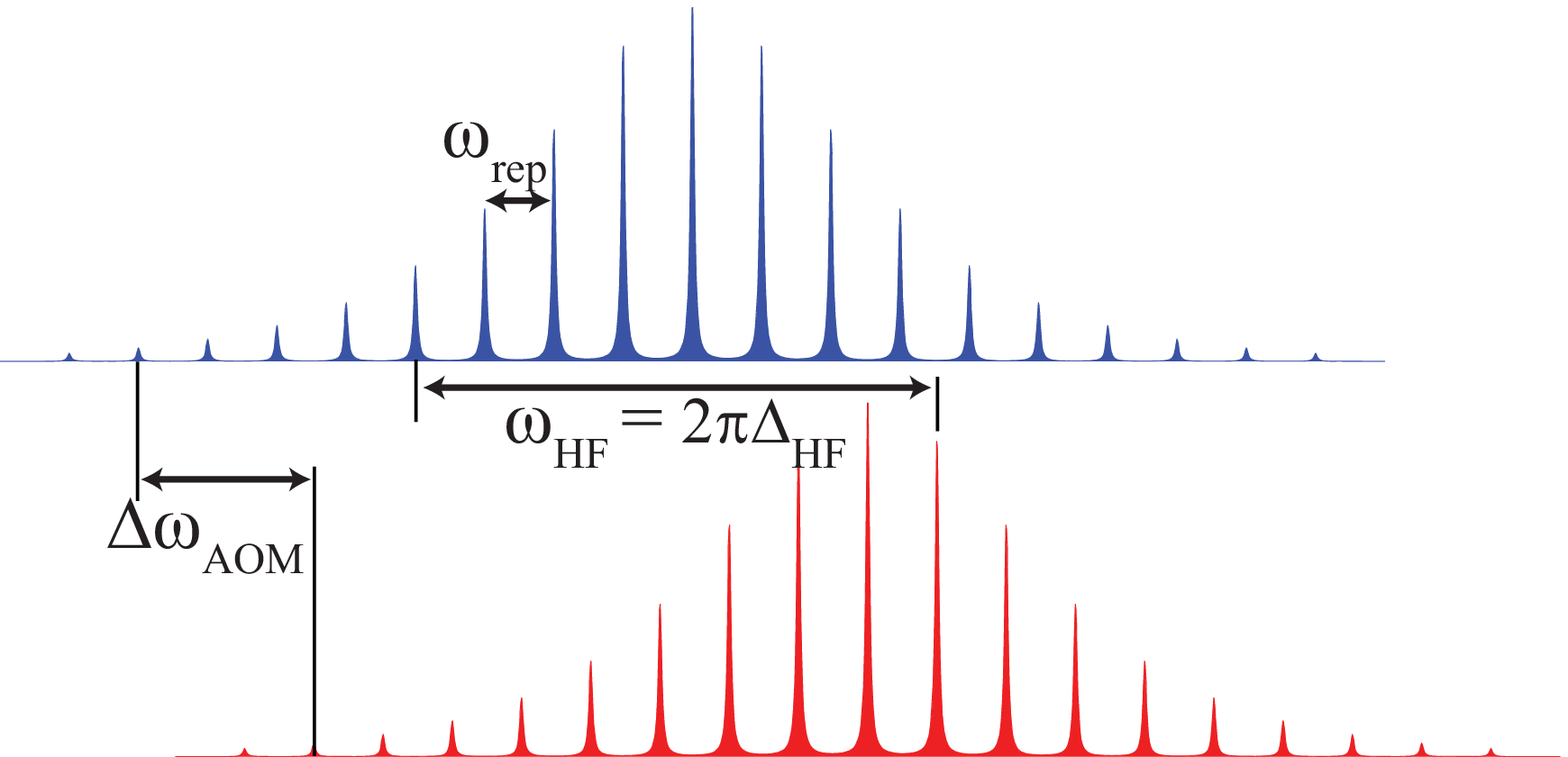}
		\label{fig:ramanComb}}
		\quad
	\subfigure[]{
		\includegraphics[trim=0cm 0cm 0cm 0cm, width=.3\textwidth] 
		{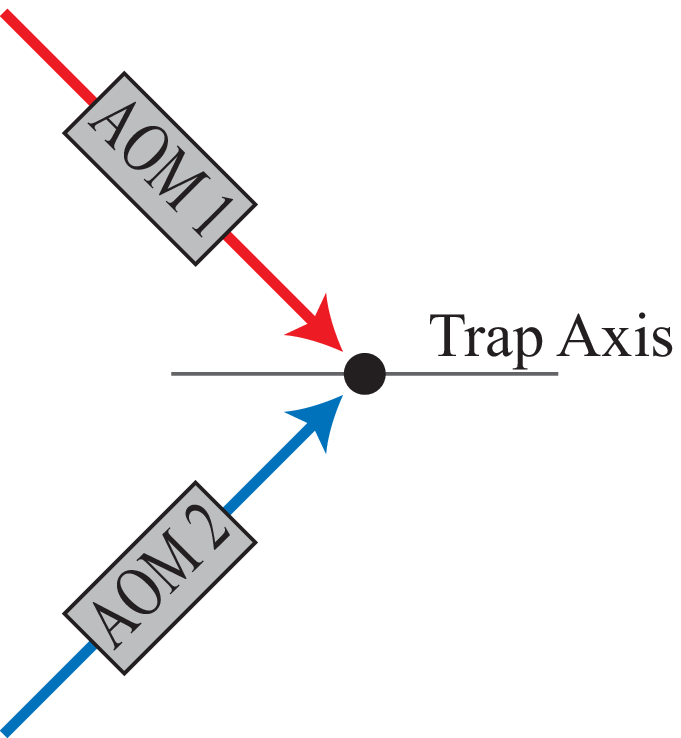}
		\label{fig:nonCoPropRaman}}
\caption{(color online)
(a) Schematic of mode-locked titanium sapphire laser repetition rate ($\sim$\,76\,MHz) stabilization. To lock the repetition rate a small portion of the laser output is incident on a photodiode (PD) whose signal is routed through a tunable bandpass (BP) filter (bandwidth 40\,MHz) that selects the 166$^{th}$ harmonic of the laser repetition rate. This signal is mixed with a reference oscillator (LO) and sent to a proportional-integral-differential (PID) regulator which provides the feedback to the laser cavity length through a high voltage amplifier (HV Amp).   (b) Diagram of frequency combs where $\omega_{\mathrm{rep}}$ is the stabilized laser repetition rate,  $\Delta\omega_{\mathrm{AOM}}$ is the difference frequency of the two combs, and $\omega_{\mathrm{HF}}=\Delta\omega_{\mathrm{AOM}}+n*\omega_{\mathrm{rep}}$, with $n\,=\,166$. All comb teeth pairs with a difference frequency resonant with $\omega_{\mathrm{HF}}$ contribute to driving the transition.  (c) The two non-co-propagating frequency combs yield a $\Delta$k perpendicular to the trap axis.}
\label{fig:pulsedSetup}
\end{figure}

The qubit states used for these experiments are the \sfzc and \sfoc hyperfine ground states (denoted \ket{0} and \ket{1} respectively) of the \yb ion. These ``clock" states are chosen for their  insensitivity to magnetic field fluctuations (to first order) enabling long coherence times (figure \ref{fig:RamanQubit}). Manipulation and detection of the \yb hyperfine qubit have been well documented in reference \cite{Olmschenk2007}.  Coherent transfer between qubit states can be accomplished using a resonant microwave field at $\Delta_{\mathrm{HF}}$\,=\,12.6\,GHz, or by stimulated  Raman transition using two laser beams with a frequency difference of $\Delta_{\mathrm{HF}}$ (figure \ref{fig:RamanQubit}).  

The electronic state of the \yb hyperfine qubit is measured by applying near-resonant light ($\sim$\,100\,mW/cm$^2$) for the \sfo $\rightarrow$ \pfz transition of the ion \cite{Olmschenk2007}. The beam excites \sfo $\rightarrow$ \pfz transitions while the \sfz state remains unchanged. The \pfz state decays back down to one of the \sfo Zeeman states and the excitation can be repeated many times.  The transition between the \sfz state and the \pfz state is forbidden by atomic selection rules, leaving the \sfz state dark throughout the detection process.  For a given polarization of the excitation beam, linear superposition states of the \sfo sublevels that are decoupled from the pump beam, known as coherent dark states \cite{Berkeland2002} are formed.  To prevent the ion from remaining in such dark states, a magnetic field of 3-5 Gauss is provided by wire coils located outside the vacuum chamber to lift the degeneracy of the \sfo states and thus ``destabilize" the coherent dark states over time \cite{Mahdifar2008}.  Due to the small possibility of the ion going from the \pfz state to the \dfo state (not shown in figure \ref{fig:RamanQubit}) a 935\,nm laser is used to pump the ion back into the \pfz $\leftrightarrow$ \sfo manifold. Sidebands are placed on the 935\,nm laser to ensure any population in the $^2$D$_{3/2}$ state sublevels (not shown in figure \ref{fig:RamanQubit}) is also pumped out.  This scheme results in ion fluorescence if the ion is in the \sfo state, while the ion remains dark if it is in the \sfz state.

Continuous wave (CW) external cavity diode lasers are used to cool, initialize, and detect the state of the qubit. The diode laser addressing the main transition (369.5\,nm) is frequency locked to a transfer cavity resonant at both 369.5\,nm and 780\,nm wavelengths. The transfer cavity length is locked to a 780\,nm diode laser which is frequency stabilized to a rubidium vapor cell using saturated absorption spectroscopy. 

State initialization is performed by adding a 2.1\,GHz sideband onto the cooling beam via a resonant electro-optic modulator (EOM).  After several scattering events the ion is pumped into the \ket{0} state \cite{Olmschenk2007}. For Doppler cooling, a 14.7\,GHz blue sideband on the cooling beam is needed to pump the ion from the dark \sfz state to the \pfo state.  This is accomplished using the second order sideband produced by a bulk resonant EOM driven at 7.37\,GHz.

Frequency combs from phase-locked ultrafast lasers are well suited for driving Raman transitions between the hyperfine qubit states due to their ability to span large frequency gaps while providing very fine frequency tuning \cite{Hayes2010}.   For our experiments a picosecond titanium-sapphire laser with a center frequency near 739\,nm and 76\,MHz repetition rate ($\omega_{\mathrm{rep}}/2\pi$) is used. The output is frequency doubled to yield a central frequency near 369.5\,nm, red-detuned  by $ \delta$\,=\,395\,GHz  from the ion resonance. The repetition rate of the pulsed laser is stabilized to ensure that the frequency difference between the two Raman beams plus an integer multiple of the repetition rate spans $\Delta_{\mathrm{HF}}$.

The repetition rate of the mode-locked laser is stabilized by monitoring the beat frequency between the 166$^{th}$ harmonic of the laser repetition rate with a stable oscillator near $\Delta_{\mathrm{HF}}$, and controlling the cavity length with the error signal (figure \ref{fig:lockingfig}).  The Raman transition is driven by first splitting the frequency comb into two beams, then shifting the frequency of each beam using acousto-optic modulators (AOMs) (figure \ref{fig:nonCoPropRaman}).

To manipulate the qubit states without affecting the ion's motion, co-propagating Raman beams can be used.  The two combs can be combined in a single beam by driving a single AOM with two modulation frequencies.  For qubit manipulation that involves changes in the ion's motional degree of freedom, we use two beams with a finite k-vector difference $\Delta$k\,=\,k$_1$\,-\,k$_2$ to impart a change in the ion's momentum. The two beams are aligned such that $\Delta$k lies perpendicular to the trap RF axis, allowing for addressing of the radial modes of motion within the trap.

\section{Results}
\subsection{Trapping and cooling}

To load ions into the trap, an oven with solid metallic Yb is resistively heated to provide a beam of neutral Yb atoms to the trapping volume. A 399\,nm diode laser (intensity $\geq$\,250\,W/cm$^{2}$) along with the 370\,nm cooling laser is used to resonantly photoionize the neutral atoms. Typically the ovens are heated for no longer than five minutes to trap a single ion.  Once the atom is ionized in the trapping volume it is immediately cooled with Doppler cooling light ($\sim$\,325\,mW/cm$^2$) detuned 20\,-\,30\,MHz from the \sfo $\rightarrow$ \pfz resonance with constant 935\,nm re-pumping.  In this system ion lifetimes of greater than 10 hours are routinely observed while performing experiments, provided Doppler cooling is applied with a duty cycle of greater than 20\%.  Without cooling, ion lifetimes as long as 20 minutes are observed.

\subsection{Trap characterization}
\subsubsection{Trap frequencies}

The trap secular frequencies are measured by applying an additional excitation signal on the trap RF voltage, accomplished using an RF power combiner.  A motional mode is resonantly excited when the resulting sideband frequency approaches one of the trap secular frequencies ($\omega_{\mathrm{trap}}$), or their harmonics. With the detection beam red-detuned by several atomic linewidths, the excitation of a motional mode causes an increase in the photon scatter rate. The longitudinal ($y$-axis) mode, parallel to the RF rails, is not found using this method implying that the trap can be well approximated as linear.  Radial trap frequencies of 1.48\,MHz and 2.10\,MHz were found using this method. These frequencies are consistent with simulated values corresponding to an applied RF voltage of 220\,V, which is well under the 300\,V maximum voltage tolerated by the trap.

\subsubsection{Micromotion}

Micromotion of the ion is a result of the misalignment of the RF null and DC nulls in the $x$-$z$ plane (as shown in figure \ref{fig:tbirdCrossSection}), where the ion's motion is driven at the RF frequency by the residual RF field at null defined by the RF and DC fields. Elimination of the micromotion is critical for the realization of multi-qubit gates. There are three detection mechanisms used to determine if the nulls are overlapping. The first method is to monitor  the lineshape of the transition between the \sfo and \pfz states. The presence of micromotion causes a spread in the velocity of the ion, which results in a Doppler broadening of its natural Lorentzian lineshape.  This method provides a rough measure of the offset between the RF and DC nulls. 

A more sensitive measure of micromotion correlates the  time of arrival of the scattered photons with the phase of the applied RF voltage\cite{Berkeland1998}. An ion located away from the RF null experiences an oscillating force at the RF frequency. For this method, the cooling laser is tuned roughly one linewidth away from the atomic resonance, where the scattering rate depends strongly on the detuning so that the ion brightness difference between when the ion sees a red or blue Doppler shift is maximized.   Since the ion is more likely to scatter a photon when it is moving towards the beam (blue shifted) than away (red shifted), we see a modulation in the photon arrival time measured with respect to the RF phase.  The measure that quantifies micromotion is the contrast of this time-correlated photon arrival fringe.  This method is only sensitive to micromotion along the laser propagation direction.  We are unable to detect micromotion along the $z$-axis (perpendicular to the electrode surfaces) using this method.

\begin{figure}[ht]
\centering
	\subfigure[]{
		\includegraphics[trim=0cm 0cm 0cm 0cm, width=.45\textwidth]
		{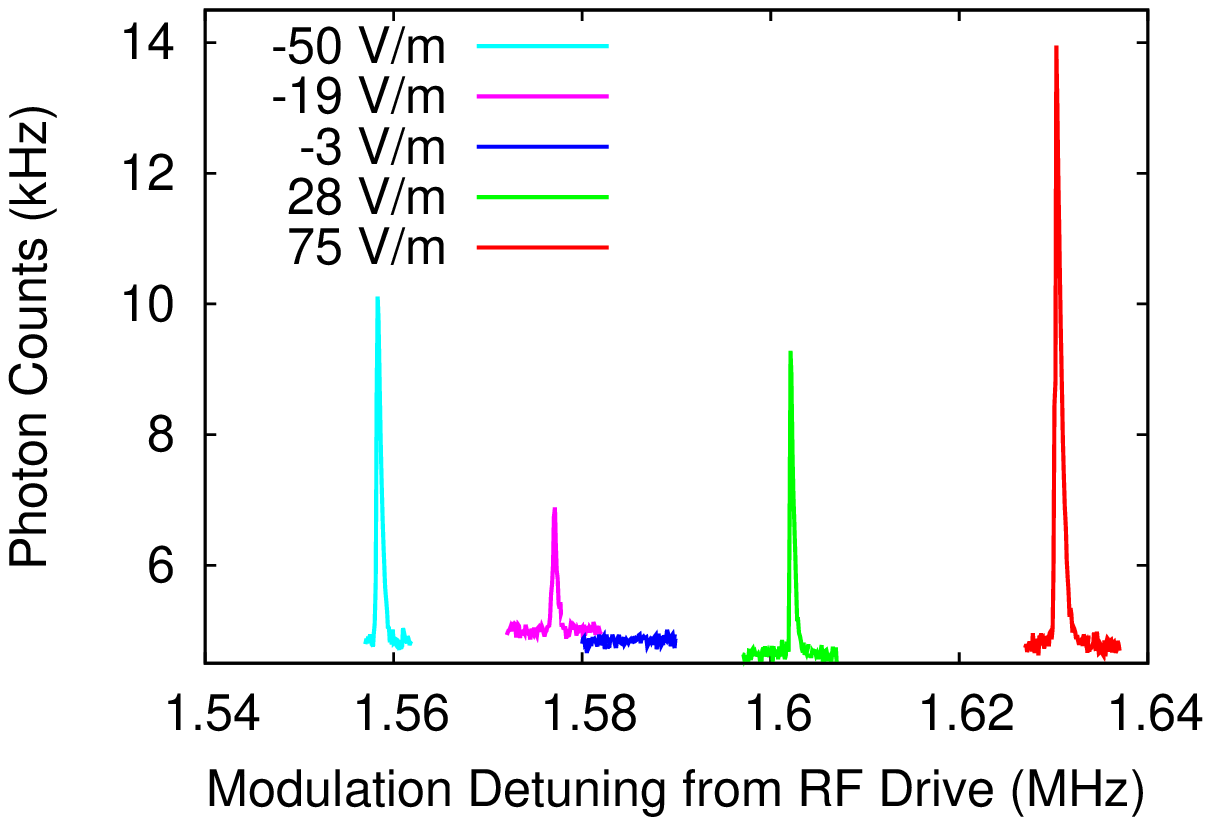}
		\label{fig:micromotion0}}
		\quad
	\subfigure[]{
		\includegraphics[trim=0cm 0cm 0cm 0cm, width=.45\textwidth]
		{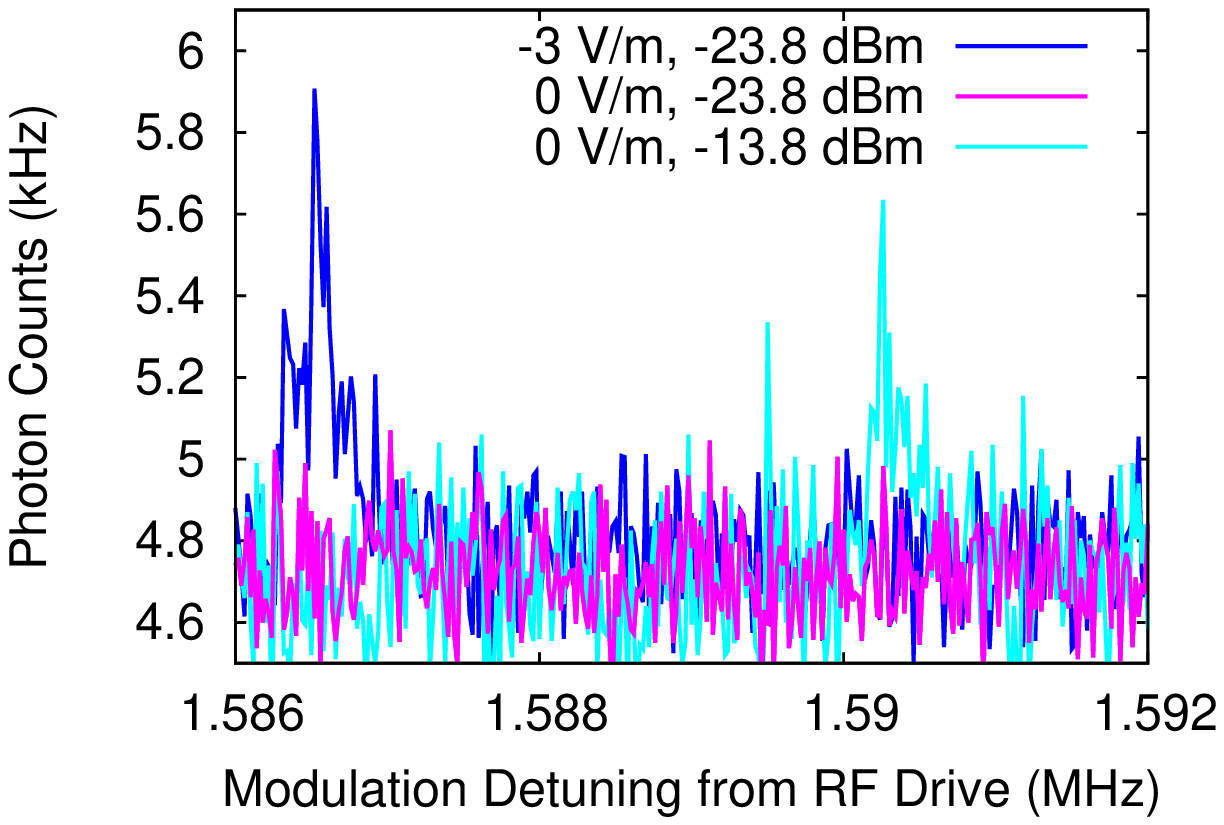}
		\label{fig:micromotion1}}
		\quad
\caption{(color online) Ion brightness as a function of the frequency detuning of the added excitation signal from the RF signal.  Each line corresponds to a finite residual electric field at the RF null, as determined from simulation.  As the micromotion is minimized the observed increase in brightness when the frequency of the modulation force approaches the trap frequency is also minimized.  (a) Data taken for various settings of the DC potential to reduce the stray field at the RF null with an excitation modulation amplitude of -33.8\,dBm applied through a helical resonator ($\omega_{\mathrm{RF}}/2\pi\,=\,27.8$\,MHz, $Q_{\mathrm{loaded}}$\,=\,280) and a trap depth of $\sim$\,86\,meV.  (b)  As the micromotion is minimized the amplitude of the added excitation signal must be increased to achieve a similar change in ion brightness.}
\label{fig:micromotion}
\end{figure}

To detect micromotion along the $z$-axis, a third method is used to resonantly excite the micromotion \cite{Ibaraki2011}.  Similar to the setup used to determine trap frequencies, we add an additional excitation signal onto the RF signal applied to the trap.  The mixing of the excitation signal and the RF signal at the trap results in an extra modulation force on the ion at the difference frequency of the two.  If the ion is positioned slightly off the RF null it will experience this force exciting the ion's motional degree of freedom, while it will remain unaffected if located at the RF null. When a red-detuned cooling beam is used, the resonant excitation is translated into ion brightness.  By sweeping the frequency of the  excitation signal, we see an increase in the ion brightness when the frequency of the added excitation signal approaches $\omega_{\mathrm{RF}}$\,$\pm$\,$\omega_{\mathrm{trap}}$.  As we change the DC voltages to bring the DC null closer to the RF null, the micromotion is reduced and the ion brightness enhancement is reduced (figure \ref{fig:micromotion0}).  Under reduced micromotion amplitude, the excitation signal power must be increased to see a similar change in brightness (figure \ref{fig:micromotion1}).

Using these methods we were able to fully compensate for micromotion only in the trap containing capacitors on all DC electrodes. For traps without on-package filter capacitors, residual RF voltage is coupled to the DC electrodes. If the RF signal pickup at the DC electrodes involves a finite phase shift, residual RF voltage appears at the RF null and the micromotion cannot be fully compensated.  All DC electrodes must therefore be filtered to eliminate any residual RF pickup.

\subsubsection{Heating rate}

It is generally observed that ions in surface traps are more susceptible to an increase in motional quanta, commonly referred to as ion heating, when compared to their macroscopic counterparts \cite{Deslauriers2004, Hite2012}.  Uncontrolled changes in the motional quanta  will degrade the fidelity of multi-qubit gates that utilize the motional degree of freedom, and must be minimized.  Compared to macroscopic traps, the ions are typically trapped much closer to the trap electrodes in surface traps, increasing the effect of fluctuating fields from the trap electrodes on the ion motion.  Additionally, trap frequencies in microfabricated traps tend to be lower since they cannot typically tolerate the high RF voltages used on macroscopic traps, leading to higher heating rates when subject to the same heating source.  A sensitive measurement of the ion heating rate can be performed by cooling the ion close to the ground state of motion, allowing the ion to heat for a controlled amount of time, and then measuring the average number of motional quanta contained in the system.  The heating rate of the trap containing capacitors on all of the DC electrodes was measured using this approach.

The ion is brought close to the motional ground state by Raman sideband cooling using optical frequency combs, as described in section \ref{sec:qubit} \cite{Diedrich1989}.  The optical frequency comb produced by the repetition-rate-stabilized pulsed laser is divided into two beams, and the frequency is shifted in each beam by an AOM. The difference frequency between the driving signals for the AOMs is tuned such that pairs of comb teeth, one from each beam, exactly match the energy difference between the \ket{0} and \ket{1} states ($\Delta_{\mathrm{HF}}$) plus or minus the trap frequency along an axis on which  $\Delta$k has a projection (see figure \ref{fig:nonCoPropRaman}). Each beam has an average intensity of 335\,W/cm$^{2}$ with a detuning of 395\,GHz from resonance.

To perform Raman sideband cooling, first 2.1\,GHz sidebands are added onto the red detuned Doppler cooling light for a time sufficient to pump the ion to the \ket{0} state ($\sim$\,20\,$\mu$s). Then, the frequency combs are turned on, tuned to the first red motional sideband of the \sfzn to \sfonm transition for a $\pi$-time which has a finite probability to subtract a motional quantum from the system. The ensuing pumping that places the ion back in the \ket{0} state on average leaves the motional state unchanged.  This sequence (motional transition, pump to \ket{0}) is repeated many times on both transverse motional modes (1.5 and 2.1\,MHz) to condense the motional distribution near the ground state.

\begin{figure}[ht]
\centering
	\subfigure[]{
		\includegraphics[trim=0cm 0cm 0cm 0cm, width=.45\textwidth]
		{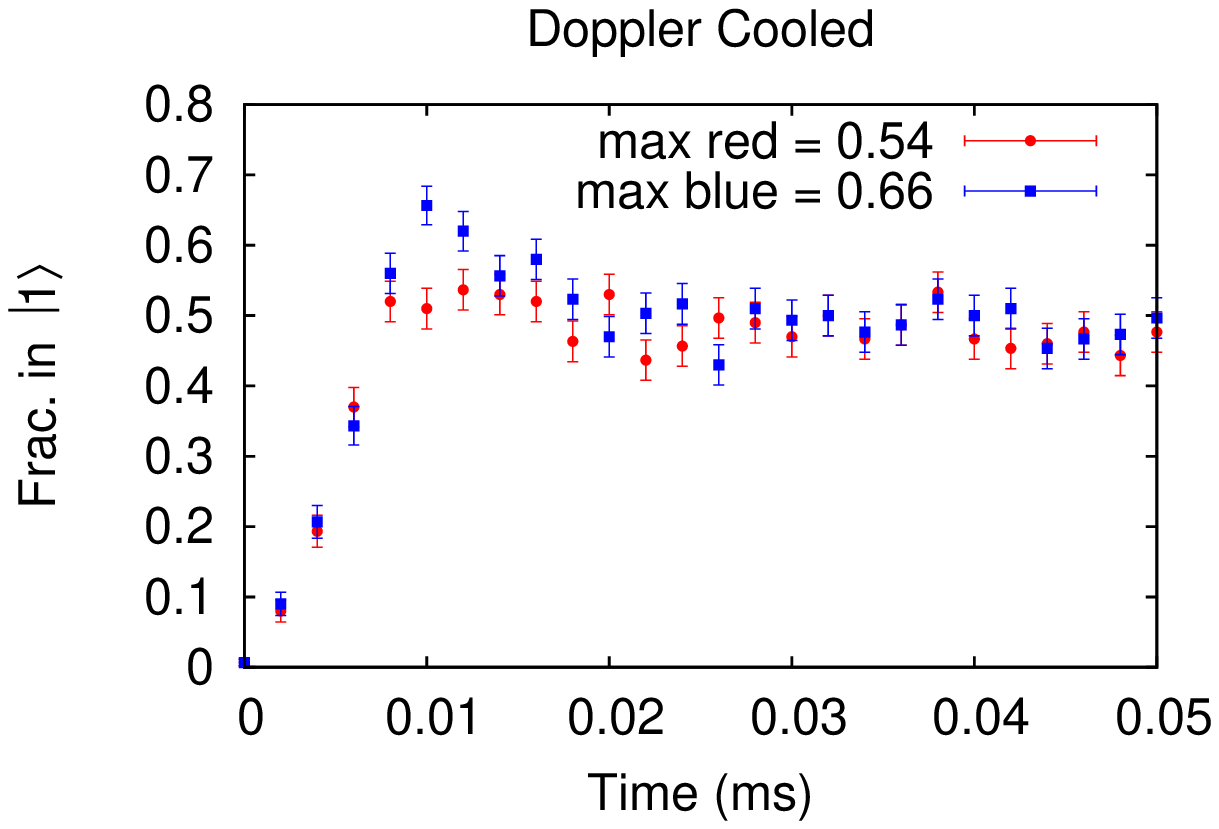}
		\label{fig:ramanRabiDoppler}}
		\quad
	\subfigure[]{
		\includegraphics[trim=0cm 0cm 0cm 0cm, width=.45\textwidth] 
		{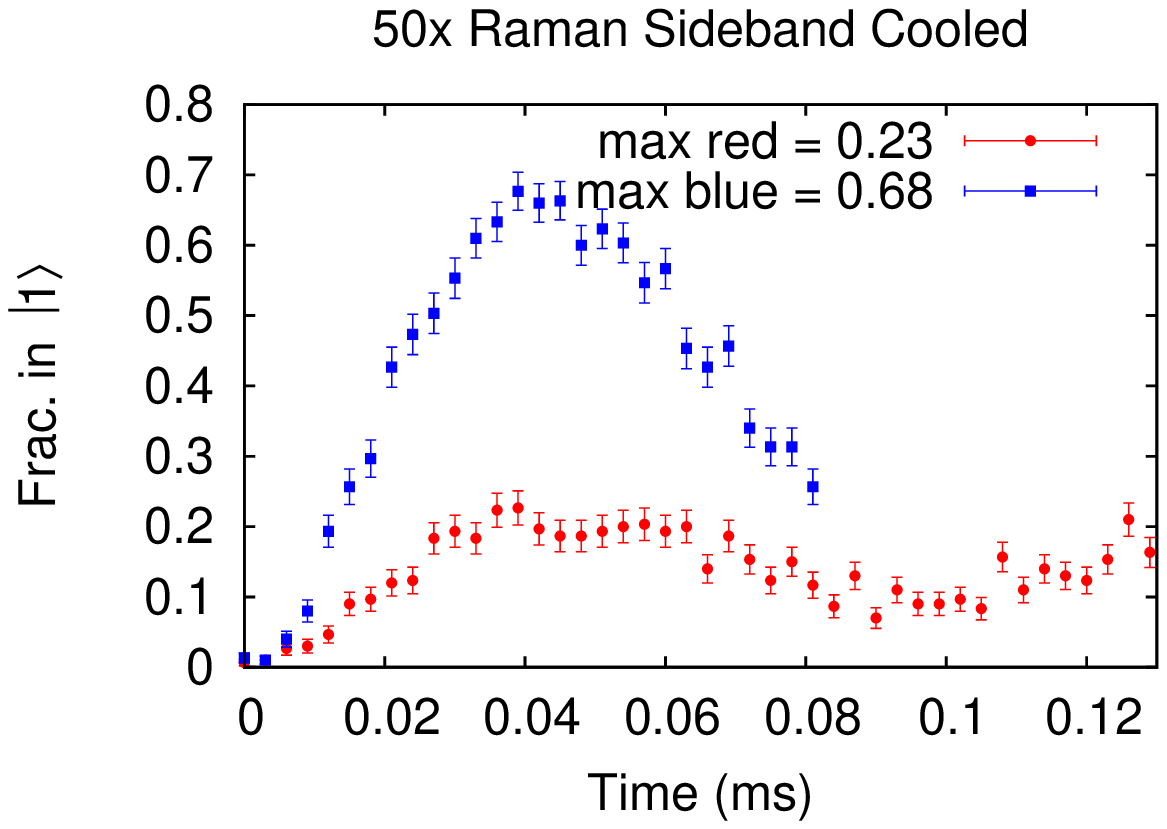}
		\label{fig:ramanRabiCool}}
		\quad
	\subfigure[]{
		\includegraphics[trim=0cm 0cm 0cm 0cm, width=.45\textwidth] 
		{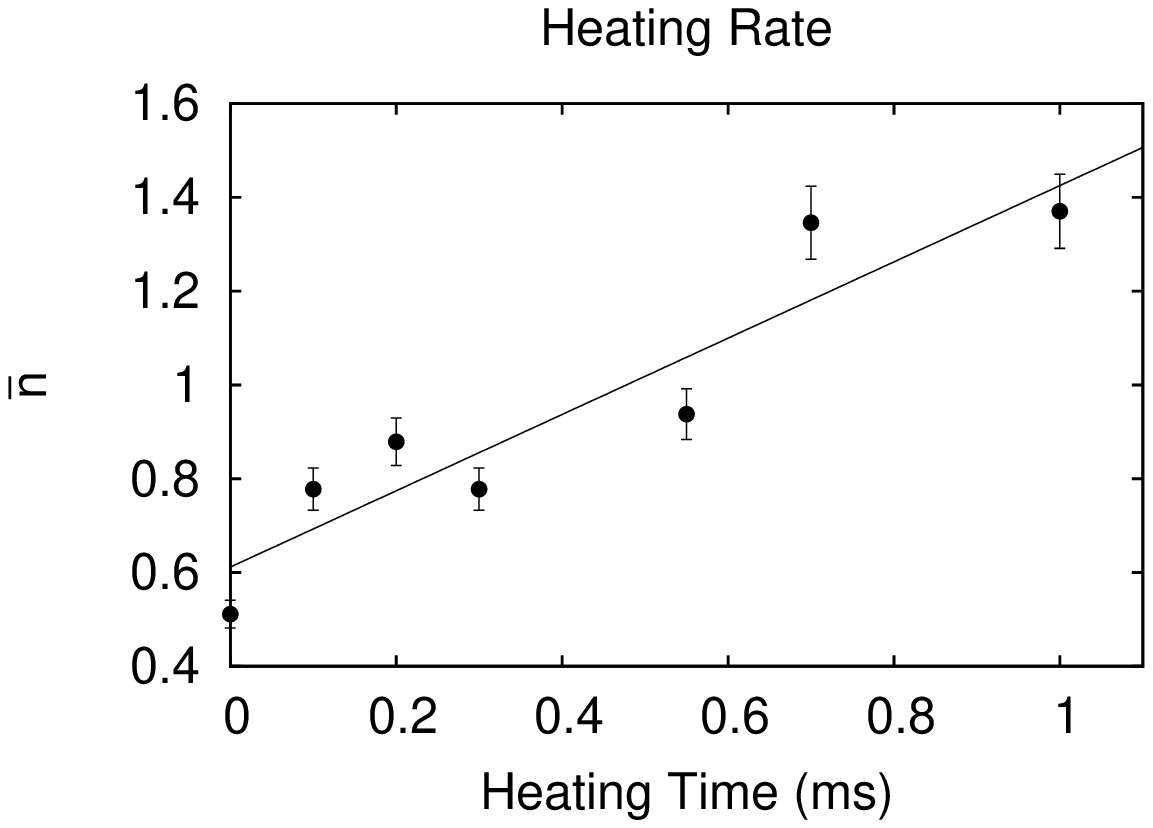}
		\label{fig:heating2}}
\caption{
(a) Rabi flopping on the first blue and red motional sidebands after only Doppler cooling, $\bar{n}$\,=\,4.5. (b) Rabi flopping on the first blue and red higher frequency transverse motional sidebands after 50 iterations of Raman sideband cooling on both transverse motional modes, $\bar{n}$\,=\,0.5.  Error bars represent standard error for the mean of a binomial distribution. (c) After 50 iterations of Raman sideband cooling on both transverse motional modes the ion is allowed to heat.  Comparison of the maximum brightness of the Rabi cycle on the first red and blue sideband yield an average number of motional quanta.  A linear fit reveals a heating rate of 0.8\,$\pm$\,0.1 quanta/ms.}
\label{fig:figure4}
\end{figure}

The average number of motional quanta in the system at a given time is determined by examining the motional sideband amplitudes.  This is accomplished by examining the first Rabi cycle on both the first red and first blue motional sideband, where the ion, initially in the dark (\ket{0}) state, is driven to the bright state (\ket{1}) by subtracting and adding a motional quantum, respectively (figure \ref{fig:ramanRabiCool}).  The maximum probability of measuring the ion to be in the bright ($\ket{1}$) state is equivalent to the amplitude of that motional sideband. The maximum probability of bright state measurement during Rabi flopping for the first red and blue motional sidebands can be used to calculate the average number of motional quanta in the trap potential by \cite{Monroe1995}:

\begin{equation}
\bar{n} = \frac{A_{red}}{A_{blue}-A_{red}},
\end{equation}
where $A_{red}$ and $A_{blue}$ are the maximum amplitude of a Rabi cycle of the red (\ket{n} to \ket{n-1}) and blue (\ket{n} to \ket{n+1}) motional sideband respectively. 

Measurements were performed on the higher frequency radial motional mode (2.1\,MHz), with an RF drive frequency $\omega_{\mathrm{RF}}/2\pi\,=\,27.8$\,MHz.  After Doppler cooling we observe $\bar{n}$\,=\,4.5 (figure \ref{fig:ramanRabiDoppler}).  Raman sideband cooling is used to cool the ion below the Doppler limit. The average number of motional quanta present after 50 iterations of Raman sideband cooling is $\bar{n}$\,=\,0.5 (see figure \ref{fig:ramanRabiCool}). To measure the heating rate, the ion is placed in the dark for up to 1 ms after Raman sideband cooling, then the Rabi oscillations for both the blue and red sideband are obtained to estimate $\bar{n}$ after the ion has heated (figure \ref{fig:heating2}).

The electric-field noise power spectral density ($S_E(\omega)$) is typically used to compare the ion heating rate between different traps and ion species \cite{Hite2011, Deslauriers2004, Turchette2000}, and is given by

\begin{equation}
S_E(\omega)=\frac{4m\hbar\omega}{q^2}\dot{\bar{n}}.
\end{equation}
The electric-field noise power spectral density is normalized by the ion mass $m$ and charge $q$, and the frequency of the motional mode ($\omega$) under investigation.  The integrated electric-field noise power for this system was found to be $\omega$$S_E(\omega)$\,=\,6.3$\times$10$^{-4}$\,V$^2$/m$^2$ corresponding to $\dot{\bar{n}}$\,=\,0.8\,$\pm$\,0.1 quanta/ms, shown in figure \ref{fig:heating2}.  This value is significantly lower than previous measurements by Allcock et al. in a nearly identical ion trap \cite{Allcock2012}, and is consistent with measurements obtained from various non-cryogenic ion traps \cite{Hite2011}.  The lowest heating rate was observed in the trap with a gold coated surface which contained {\em in situ} capacitive filters that enabled complete compensation of micromotion.  For this trap, the NEG pump was placed a few millimeters away from the trap surface.  The proximity of the pump to the trap may have assisted in decreasing the ion collision rate with background gas molecules, an important factor in achieving exceptionally long trapping times.

\subsection{State detection}

\begin{figure}
\centering
	\includegraphics[trim=0cm 0cm 0cm 0cm, width=.5\textwidth]{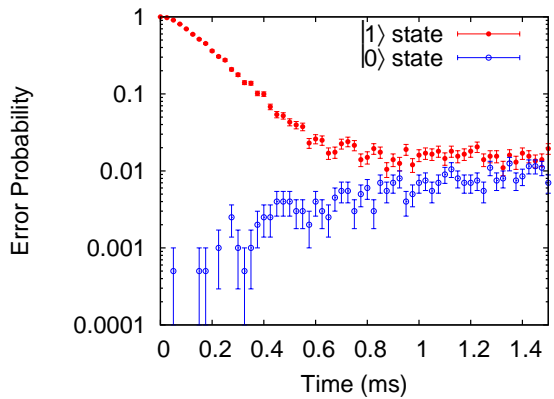}
\caption{Error probability of the state detection for the \ket{0} and \ket{1} states using a threshold of 1.5 photons where detection of 0 or 1 photons is noted as a \ket{0} state and detection of 2 or more photons results in \ket{1} state measurement.  An average state detection fidelity of 98.8(0.3)\% is achieved in 1\,ms.}
\label{fig:stateDetect}
\end{figure}

High fidelity state detection is crucial for quantum information processing experiments. Here we employ standard state-dependent fluorescence detection. The ion fluorescence is collected using a custom objective with a numerical aperture of 0.26, and is directed through a bandpass filter (Semrock FF01-370/6, 6\,nm width) to a photo-multiplier tube (Hamamatsu 10682-210). Ideally no photons are scattered when the  ion is in the \ket{0} state while many photons are scattered from an ion in the \ket{1} state. In reality, there are detector dark counts and background counts arising from detection light scattered off the trap surface that are collected. By analyzing the distributions of the number of photons during the detection interval for the bright and dark states, a threshold of 1.5 photons is determined. If the PMT collects 0 or 1 photon counts during the detection interval, the ion is determined to be in the \ket{0} state. If the PMT collects more than 1 photon count during the detection interval, the ion is then determined to be in the \ket{1} state. This yields a best detection fidelity of 98.8\,$\pm$\,0.3\% for a detection interval of 1 ms (see figure \ref{fig:stateDetect}).

The largest source of error for detection of the \ket{1} state  can be attributed to off-resonant excitation of the \ket{1} state to the \pfo state during the detection process (2.1\,GHz detuned), which can decay to the \ket{0} state and quench the fluorescence.  The state detection fidelity could be improved  by increasing the photon collection and detection efficiency, and decreasing the detector dark counts and background counts \cite{Olmschenk2007, Noek2013}.

\subsection{Microwave gates}

\begin{figure}[ht]
\centering
	\subfigure[]{
		\includegraphics[trim=0cm 0cm 0cm 0cm, width=.5\textwidth]
		{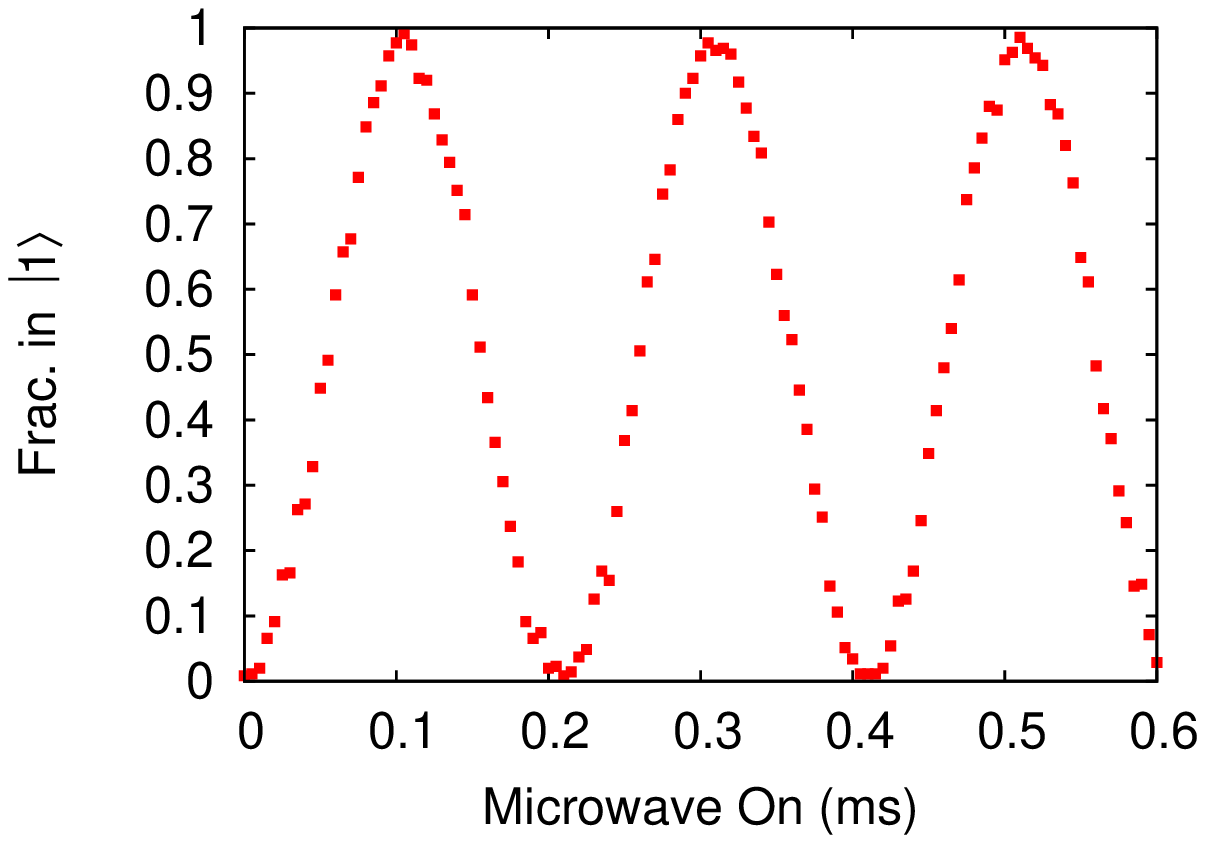}
		\label{fig:microwave}}
		\quad
	\subfigure[]{
		\includegraphics[trim=0cm 0cm 0cm 0cm, width=.45\textwidth] 
		{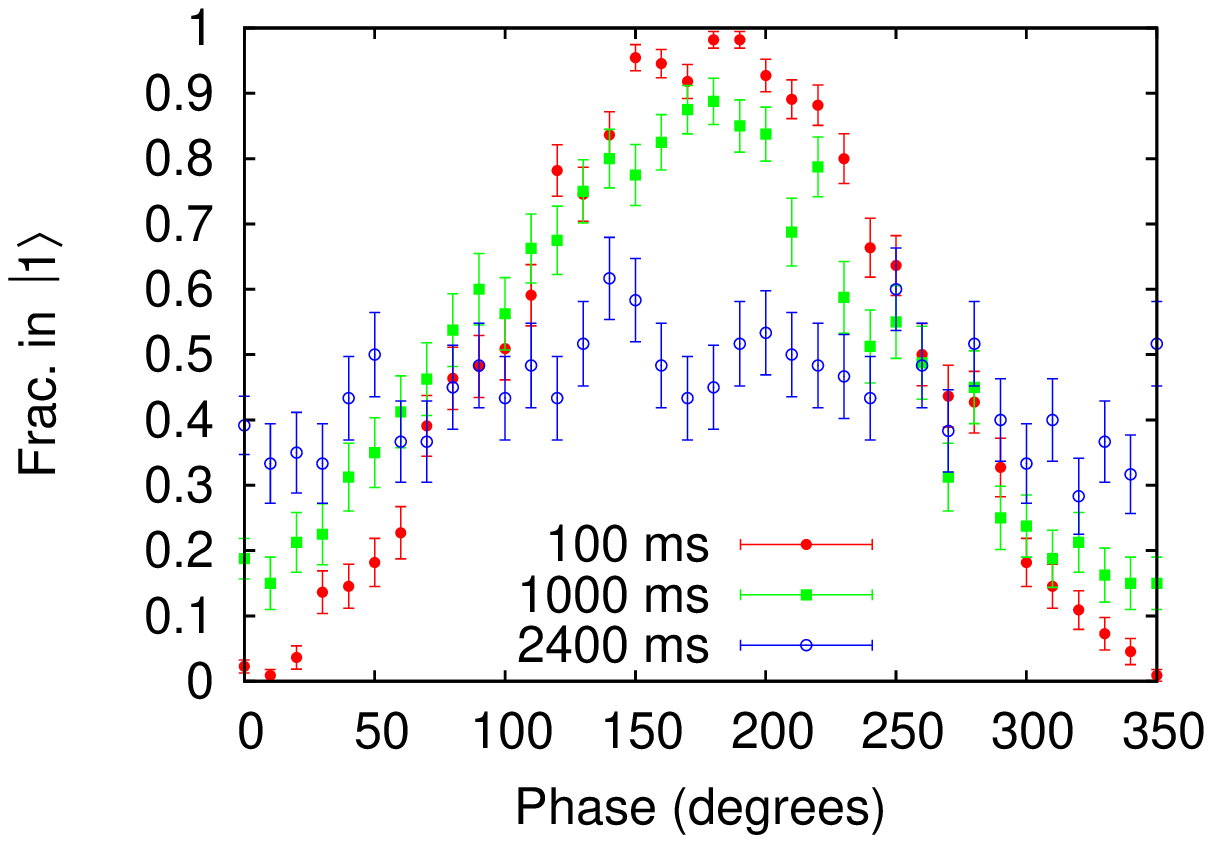}
		\label{fig:threeRamFringe}}
		\quad
	\subfigure[]{
		\includegraphics[trim=0cm 0cm 0cm 0cm, width=.45\textwidth] 
		{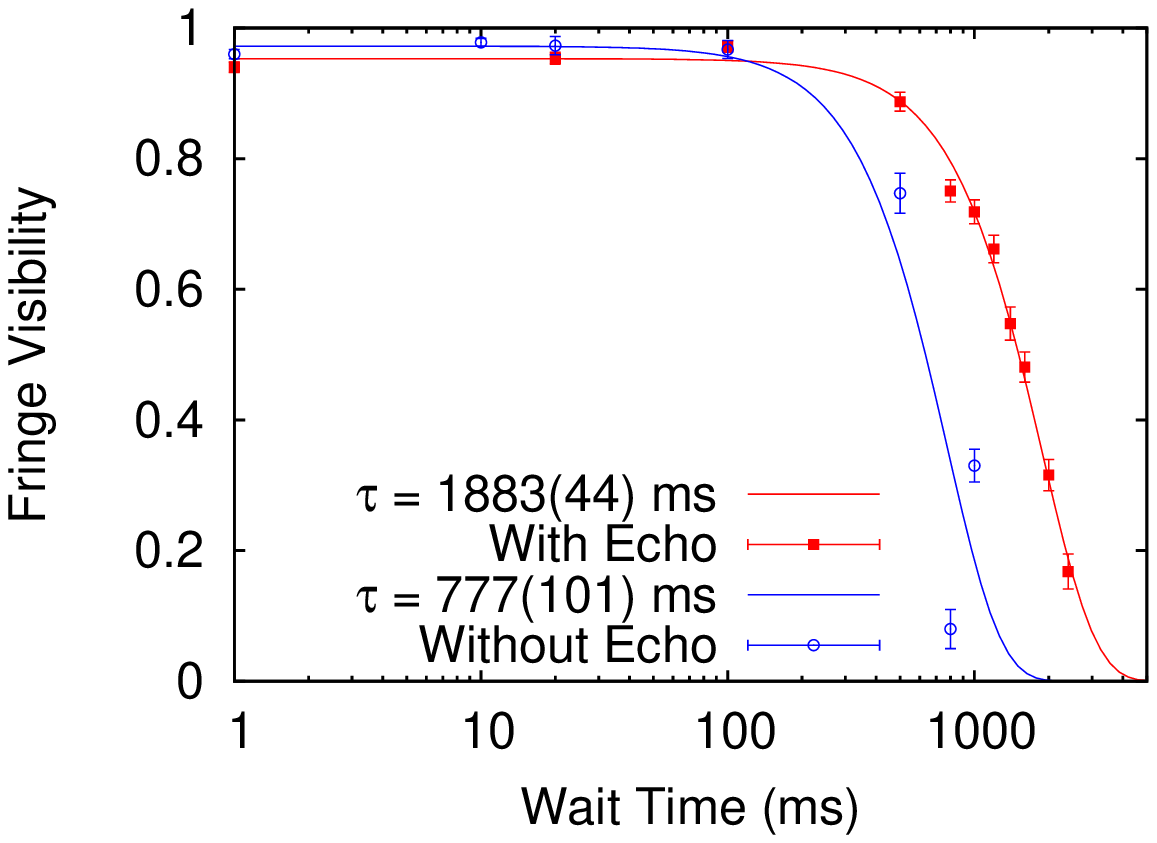}
		\label{fig:microRamLog}}
\caption{
(a) The fraction of experiments in which the qubit was found in the \ket{1} state as a function of the duration of the applied microwave field.  (b) Example Ramsey fringes for 100, 1000, and 2400\,ms wait time between the $\pi$/2 pulses. (c) Microwave Ramsey results, with and without a single spin-echo pulse. Each data point is extracted from 40\,-\,100 experiments.}
\label{fig:microwave}
\end{figure}

By addressing the $\Delta_{\mathrm{HF}}$\,=\,12.6\,GHz energy splitting between the qubit states directly with a resonant microwave field, we can perform single qubit rotational gates on the hyperfine qubit. This is accomplished by aligning  a microwave horn to the ion from outside the vacuum chamber. 

After Doppler cooling, the ion is initialized to the \ket{0} state. A 12.6\,GHz microwave field resonant with the \ket{0} to \ket{1} transition is then applied for a variable amount of time, followed by state detection. The probability to find the qubit in the \ket{1} state as a function of the duration of the applied microwave field is shown in figure \ref{fig:microwave}. After the microwave field is applied for $\sim$\,0.1\,ms the state of the qubit is fully transferred from \ket{0} to \ket{1}.  More efficient microwave coupling can be accomplished using trap chips with integrated microwave waveguides \cite{Allcock2013}.

The qubit-microwave coherence time is characterized by a Ramsey experiment. In this experiment the ion is first placed in a superposition state of \ket{0} and \ket{1} by a $\pi$/2 pulse ($\sim$\,50\,$\mu$s). This pulse is followed by a time delay, then another $\pi$/2 pulse with an adjustable phase. The probability to find the qubit in \ket{1} state as a function of the phase of the second pulse results in a fringe (figure \ref{fig:threeRamFringe}). The fringe visibility as a function of time delay is fit to a Gaussian function to deduce a coherence time of 780\,$\pm$\,101\,ms (figure \ref{fig:microRamLog}).

To remove dephasing errors arising from a discrepancy between the ion clock and the local microwave oscillator, a sequence of three microwave pulses can be used, provided this dephasing occurs at a rate slower than the time required for one experiment \cite{Andersen2003, Langer2005}.  Applying this spin-echo technique increases the measured coherence time to 1883\,$\pm$\,44\,ms.

\subsection{Raman gates}

A Ramsey experiment was also performed using a single-path pulsed laser beam modulated by an AOM at two frequencies with an average intensity of 523\,W/cm$^{2}$  at the ion. The beam has $\sigma$ polarization such that the transition, as shown in figure \ref{fig:RamanQubit} (light and dark red and blue)  is accomplished. A Ramsey experiment yields a qubit coherence time of 1433\,$\pm$\,37\,ms with a single spin-echo pulse. The reduced coherence time observed for the Raman gates in comparison with the microwave gates is attributed to the residual intensity fluctuation of the Raman beams at the ion.

\section{Conclusions}

We have demonstrated initialization, detection and manipulation of the \yb qubit in a microfabricated surface trap.  Microwave single qubit manipulation was demonstrated, with a  coherence time of 780\,$\pm$\,101\,ms and 1883\,$\pm$\,44\,ms, without and with a single spin-echo pulse, respectively. 

Qubit state transitions were also driven with optical frequency combs with a center frequency detuned from the atomic resonance by 395\,GHz. Using a single beam containing two frequency combs we have demonstrated single qubit manipulation with a qubit coherence time of 1433\,$\pm$\,37\,ms with a single spin-echo pulse. As a precursor to motional gates, two beam paths, each with a single frequency comb, were used to drive transitions between qubit and motional states. Raman sideband cooling was performed, cooling the ion from 4.5 to 0.5 average motional quanta. Ion heating rates as low as 0.8\,$\pm$\,0.1 quanta/ms have been observed at a transverse trap frequency of 2.1\,MHz. 

The quantum computing primitives demonstrated in this work show that \yb ions in microfabricated ion traps are strong candidates for scalable quantum computation.    While there is a need for demonstration of more sophisticated logic gates in these traps, the results reported here suggest that spin-motion entanglement and two-ion gates using optical frequency combs in a microfabricated surface trap are possible.

\ack
This research was funded by the Office of the Director of National Intelligence (ODNI), Intelligence Advanced Research Projects Activity (IARPA), through the Army Research Office.  Sandia National Laboratories is a multi-program laboratory managed and operated by Sandia Corporation, a wholly owned subsidiary of Lockheed Martin Corporation, for the U.S. Department of Energy's National Nuclear Security Administration under contract DE-AC04-94AL85000. SAND2011-0439P.


\section*{References}


\end{document}